\documentclass{pnastwo}

\usepackage{amssymb,amsfonts,amsmath,graphicx}

\newcommand\apj{Astrophys.~J.~}                                          
\newcommand\apjl{Astrophys.~J. Lett.~}                                          
\newcommand\apjs{Astrophys.~J. Suppl.~}                                         
\newcommand\aap{Astron. Astrophys.~}                                            
\newcommand\mnras{Mon. Not. R.~Astron. Soc.~}                                   
 
\newcommand\aj{Astron. J.~}                                            
   
\newcommand\pasp{Pub.~Astron.~Soc.~Pacific~}                
\newcommand\nat{Nat.~}

\contributor{Submitted to Proceedings
of the National Academy of Sciences of the United States of America}
\url{www.pnas.org/cgi/doi/10.1073/pnas.0709640104}
\copyrightyear{2008}
\issuedate{Issue Date}
\volume{Volume}
\issuenumber{Issue Number}

\begin{document}



\title{The formation of high-field magnetic white dwarfs from common envelopes}

\author{
J. Nordhaus\affil{1}{Department of Astrophysical Sciences, Princeton
  University, Princeton, NJ 08544,
  U.S.A.}\footnotetext{nordhaus@astro.princeton.edu},
S.~Wellons\affil{1}{},
D.~S.~Spiegel\affil{1}{},
B.~D.~Metzger\affil{1}{},
\&
E.~G.~Blackman\affil{2}{Department of Physics and Astronomy,
  University of Rochester, Rochester, NY 14627, U.S.A}}

\footlineauthor{Nordhaus et al.}

\contributor{Submitted to Proceedings of the National Academy of
  Sciences of the United States of America}

\maketitle

\begin{article}

\begin{abstract}
The origin of highly-magnetized white dwarfs has remained a mystery
since their initial discovery.  Recent observations indicate that the formation of high-field magnetic white dwarfs is
intimately related to strong binary interactions during post-main-sequence
 phases of stellar evolution.  If a low-mass companion, such
as a planet, brown dwarf, or low-mass star is engulfed by a post-main-sequence giant, gravitational torques in the envelope of the giant lead to a reduction of
the companion's orbit.  Sufficiently low-mass companions in-spiral
until they are shredded by the strong gravitational tides near the
white dwarf core.  Subsequent formation of a super-Eddington accretion disk from the
disrupted companion inside a common envelope can dramatically amplify
magnetic fields via a dynamo.  Here, we show that these disk-generated
fields are sufficiently strong to explain the observed range of magnetic field strengths for isolated, high-field magnetic
white dwarfs.  A higher-mass binary analogue may also contribute to the
origin of magnetar fields.
\end{abstract}


\abbreviations{HFMWD, high-field magnetic white dwarf; SDSS, Sloan
  Digital Sky Survey; MG, Megagauss; WD, white dwarf; CV, cataclysmic
  variable; CE, common envelope; MS, main sequence; AGB, asymptotic
  giant branch; BD, brown dwarf}

\section{Introduction}
\label{sec:Intro}
\dropcap{A} significant fraction of isolated white dwarfs form with
strong magnetic fields.  These high-field magnetic white dwarfs
(HFMWD), the majority of which were discovered via the Sloan Digital
Sky Survey (SDSS; \cite{Gansicke:2002fj,Schmidt:2003yq,
  Vanlandingham:2005kx}), comprise $\sim$10\% of all isolated white
dwarfs \cite{Liebert:2003lr}.  Surface field strengths range from a
few to slightly less than a thousand megagauss (MG) while the bulk of
the isolated WD population have measured weak fields or non-detection
upper limits of typically $\lesssim10^4-10^5$~G \cite{Kawka:2007fk,
  Valyavin:2006vn, Aznar-Cuadrado:2004qy}.

In WD-binary systems, the companion's surface is an equipotential.  In
tight binaries, this equipotential extends toward the WD, leading to
mass transfer from the companion onto the WD, a process called
``Roche-lobe overflow.''  Among systems such as these, a possibly
even larger fraction of the WD primaries are highly magnetic (i.e., $\sim$25\% of
cataclysmic variables; \cite{Wickramasinghe:2000kx}).  Magnetic
cataclysmic variables (CVs) are generally divided into two classes: AM
Herculis (polars) and DQ Herculis (intermediate polars). For reviews
on AM Her and DQ Her systems see \cite{Patterson:1994lr} and
\cite{Wickramasinghe:2000vn}.  Generally, AM Her systems are those in
which both components of the binary are synchronously rotating at the
orbital period.  In this scenario, the formation of an accretion disk
is prevented and material is funneled onto the WD via the
magnetosphere.  Polars have strong magnetic fields ($\sim$$10^7-10^8$
G), copious X-ray emission and stable pulsations in their lightcurves.
On the other hand, DQ Her systems (intermediate polars) are not
synchronously locked and have weaker magnetic fields than their polar
counterparts, often by an order of magnitude or more.  For these systems,
an accretion disk forms from which the WD is spun up.

Remarkably, {\it not a single} observed close, detached binary system
(in which the primary is a WD and the companion is a low-mass
main-sequence star) contains a HFMWD \cite{Liebert:2005uq,
  Silvestri:2006qy, Silvestri:2007fk}.  If the magnetic field
strengths of white dwarfs were independent of binary interactions,
then the observed distribution of isolated WDs should be similar to
those in detached binaries.  In particular, within 20 pc, there are 109
known WDs (21 of which have a non-degenerate companion), and SDSS has identified 149
HFMWDs ({\it none} of which has a non-degenerate companion).  Assuming binomial statistics, the maximum probability of
obtaining samples at least this different from the same underlying
population is $5.7\times10^{-10}$, suggesting at the 6.2-$\sigma$
level that the two populations are different (for details on this kind
of calculation, see Appendix B of \cite{Spiegel:2007lr}).  Furthermore, SDSS identified 1253 WD+M-dwarf binaries ({\it none} of which are magnetic).  As was
previously pointed out, this suggests that the presence or absence of binarity is crucial in influencing whether a HFMWD results
\cite{Tout:2008qy}.  These results initially seem to indicate that HFMWDs preferentially form when isolated.  However, unless there is a mechanism by which very distant companions prevent the formation of a strong magnetic field, a more natural explanation is that highly-magnetized white dwarfs became that way by engulfing (and removing) their companions. 

In the binary scenario, the progenitors of HFMWDs are those systems
that undergo a common envelope (CE) phase during post-MS evolution.  In
particular, magnetic CVs may be the progeny of common envelope systems
that almost merge but eject the envelope and
produce a close binary.  Subsequent orbital reduction via
gravitational radiation and tidal forces turn detached systems into
those which undergo Roche-lobe overflow.  For CEs in which the companion
is not massive enough to eject the envelope and leave a tight post-CE
binary, the companion is expected to merge with the core.  It was
suggested that these systems may be the progenitors of isolated HFMWDs
\cite{Tout:2008qy}.

In this paper, we calculate the magnetic fields generated
during the common envelope phase.  We focus on low-mass companions
embedded in the envelope of a post-MS giant.  During in-spiral, the
companion transfers orbital energy and angular momentum, resulting in
differential rotation inside the CE.  Coupled with convection, a
transient $\alpha-\Omega$ dynamo amplifies the magnetic field at the
interface between the convective and radiative zones where the strongest shear is available. The fields produced
from this interface dynamo are transient and unlikely to reach the
white dwarf surface with sufficient strength to explain HFMWD observations.  If, however, the
companion tidally disrupts, the resultant super-Eddington accretion disk can amplify
magnetic fields via a disk dynamo.  The disk-generated fields are
strong ($\sim10^1-10^3$ MG) and accretion provides a natural
mechanism with which to transport the fields to the WD surface.  For a range of disrupted-companion masses, the fields generated in the disk are sufficient to explain the range of observed HFMWDs.

\section{Common Envelope Evolution}
\label{sec:CE}
Common envelopes are often invoked to explain short period systems in
which one component of the binary is a compact star
\cite{Paczynski:1976fk, Iben:1993qy, Nordhaus:2006oq}.  The immersion
of a companion in a CE with a post-main sequence star can occur via
direct engulfment or through orbital decay due to tidal dissipation in
the giant's envelope \cite{Nordhaus:2010lr, Carlberg:2009lr,
  Villaver:2009fk}.  Once engulfed, the companion in-spirals due to
hydrodynamic drag until it either survives (ejects the
envelope leaving a post-CE close binary) or is destroyed.  While CEs
can form for various mass-ratio binaries, we focus on low-mass stellar and substellar
companions such that the remnant system is expected to be an isolated
white dwarf.  Since low-mass companions do not release enough orbital
energy to eject the envelope, as the orbital separation is reduced,
the differential gravitational force due to the proto-WD tidally
shreds the companion.  The disrupted companion then forms a disk
inside the CE that subsequently accretes onto the proto-WD
\cite{Reyes-Ruiz:1999lr}.  For more detail on the onset and dynamics
of the CE phase for post-MS giants and low-mass companions (i.e. low,
mass-ratio binaries), see \cite{Nordhaus:2006oq} and
\cite{Nordhaus:2010lr}.

\section{Amplification of Magnetic Fields}
\label{sec:Bfields}
As a consequence of common envelope evolution, the transfer of orbital
energy and angular momentum during in-spiral generates strong shear.
Coupled with convection, shear leads to large-scale magnetic field amplification
via an $\alpha-\Omega$ dynamo.  During in-spiral, the free energy in differential rotation
available to the dynamo is proportional to the companion mass.
In general, the more massive the companion, the more free energy in differential rotation \cite{Nordhaus:2008fk}.  The dynamo converts free energy in differential rotation into magnetic energy and therefore, strong shear leads to strong magnetic fields.

Some investigations of the dynamo in this context have imposed a
velocity field to determine what steady-state magnetic field might
arise if the velocity were steadily driven \cite{Tout:1992lr,
  Regos:1995uq, Blackman:2001sy}.  However, as the magnetic field
amplifies, differential rotation decreases.  In general, for systems
in which shear is not resupplied, a transient dynamo and decay of the
magnetic field result \cite{Blackman:2006bh}.  This scenario was
investigated as a way to generate the strong fields necessary
to power bipolar outflows in post-Asymptotic Giant Branch (post-AGB)
and planetary nebulae
(PNe) \cite{Nordhaus:2007il, Nordhaus:2008qy}.  A similar, but
weaker, dynamo (akin to the Solar dynamo) may operate in isolated Red
Giant Branch and Asymptotic Giant Branch stars \cite{Nordhaus:2008fk}.

Here, we investigate two scenarios for magnetic field generation
during a CE phase between a post-MS giant and a low-mass companion.
First, we calculate the magnetic fields at the interface between the
convective and radiative zones in the CE (hereafter referred to as the
envelope dynamo scenario) and show that this scenario, while potentially viable under special circumstances, is unlikely in general to explain the origin of HFMWDs.  As an alternative, we estimate the magnetic fields generated in the disk of a tidally disrupted companion around the proto-WD.  The disk-generated fields naturally accrete onto the WD surface and are sufficiently strong to match observations.

\subsection{Envelope Dynamo}
\label{ssec:EnvelopeDynamo}

The shear profile in a giant star that is produced by a low-mass
companion's CE-induced in-spiral leads to a dynamo in the presence of
the turbulent convective zone.
The field primarily amplifies at $R = R_{\rm conv}$, the interface between the convective and radiative
regions.  The back-reaction of field-amplification on the shear is
included with differential rotation and rotation depleting via loss of
Poynting flux and turbulent dissipation.  For the precise equations
solved, generic features of envelope dynamos, and a pictorial
representation of the dynamo geometry in isolated stellar and CE
settings see \cite{Nordhaus:2007il}.  Here we calculate an upper limit to the magnetic field that may be
generated by this mechanism, by taking the shear profile in the
stellar envelope to be Keplerian (which produces stronger shear than
could possibly be maintained in a spherical-hydrostatic star).

Before presenting details of these calculations, we note a few generic features of this type of scenario that might make the envelope dynamo an unlikely explanation of HFMWDs, irrespective of details.  There are a few reasons why a field generated at the radiative-convective boundary may not be able to produce strongly magnetized material in the vicinity of the WD.  First, as shown below, a robust upper limit to the magnitude of the envelope dynamo generated field is a few times $10^4$~G.  To explain megagauss fields at the WD surface, this requires an inwardly increasing gradient of ${\bf B}$ (e.g.~via field amplification from flux freezing), which might lead the highest-field regions to rise buoyantly, therefore never reaching the WD surface.  Second, since the envelope is transient (operates for $\sim$100 years) and the typical AGB lifetime is $\sim$$10^5$
years, the envelope would need to be ejected at the time of formation of a HFMWD.  Without fine tuning, this results in a tight, post-CE binary with an under-massive white dwarf -- the exact opposite of the observed HFMWDs.  Finally, even if the field could penetrate to the WD surface, some
mass would need to remain and/or fallback during envelope ejection to
anchor the field.  A further subtlety is that the transport of magnetic flux into the radiative layer is
likely to involve not merely isotropic diffusion but anisotropic diffusion, as a downward pumping
of magnetic flux involves anisotropic convection.  We are led to what seems to be a more
natural formation explanation, which is described in the next section.  Nevertheless, despite the aforementioned potential difficulties, we now investigate the viability of the envelope dynamo scenario. 

Our stellar evolution models were computed using the ``Evolve Zero-age
Main Sequence'' (EZ)
code \cite{Paxton:2004fj}.\footnote{http://www.astro.wisc.edu/$\sim$townsend/static.php?ref=ez-web}  We employ a zero-age main-sequence
progenitor of 3~$M_\odot$, with solar metallicity at the tip of the
AGB.  Under the assumption of Keplerian rotation, the saturated toroidal field strength at the base of the 3-$M_\odot$ progenitor's convective zone ($R_{\rm conv}\sim6\times10^{11}$~cm) is roughly $B_\phi\sim2\times10^4$~G.  To survive past the PN stage, the fields must reach the WD surface (which is the core of the AGB star, a distance $L\simeq R_{\rm conv}$ interior to the base of the convective
zone; see Fig.~3 of \cite{Nordhaus:2006oq} for a pictorial
representation), anchor, and sustain or induce a field in the WD.  The
latter scenario has been investigated by \cite{Potter:2010fk}, who
conclude that an ordered external field (potentially generated from a
dynamo) can induce a surface field on the WD that decays to a few
percent of its initial value after a million years.

For the fields to reach the WD surface,
radially inward diffusion of magnetic flux must act on a faster
timescale than magnetic buoyancy, which transports flux outward.
The buoyant rise velocity, $u$, is found by
equating the upward buoyancy force on a flux tube to the downward
viscous drag, thereby obtaining the upward terminal velocity.  It
may be represented as
$u\sim\left(3Q/8\right)\left(a/H_{\rm p}\right)^2\left(v_a^2/v\right)$,
where
$v_a = B/(4 \pi \rho)^{1/2}$
is the Alfv\'{e}n velocity,
$v$ is the convective fluid velocity, $a$ is the flux tube radius,
$H_{\rm p}$ is a pressure scale height, and $Q$ is a dimensionless
quantity of order unity \cite{Parker:1979lr}.
The time to buoyantly rise a distance $L$ from $R_{\rm conv}$, assuming
$a\sim L / 2$, is
$t_{\rm b}=L/u\sim 0.1$ years.  The diffusion timescale to traverse a
distance $L$, given a turbulent diffusivity $\beta_\phi$, is
$t_{\rm d} = L^2 / \beta_\phi$ and must be less than the buoyant-rise
time $t_{\rm b}$.  This implies a constraint on the turbulent
diffusivity: $\beta_\phi>Lu$.  For $L\sim6\times10^{11}$ cm, the buoyant-rise velocity of a
$2\times10^4$-G flux tube in an AGB star with $\rho\sim10^{-4}$ g cm$^{-3}$, requires a diffusivity
of at least a few times $10^{17}$~cm$^{2}$~s$^{-1}$; a very large and physically unlikely value.  Note however that if weaker fields are generated, the requirement on the diffusivity would lower correspondingly.  Furthermore, during one cycle half-period $\tau_{0.5}$ (defined as the time for one
reversal of the field), magnetic flux of a given sign (positive or
negative) can diffuse into the radiative layer.\footnote{Though the cycle
  period does increase in the dynamical regime, it does not deviate
  significantly from the half-cycle kinematic value we use here ($\tau\sim0.03$
  years) ; \cite{Nordhaus:2007il}).  Note that while the radiative layer is convectively stable it may still possess non-negligible turbulence.}  
  After reversal, field of the opposite sign amplifies and diffuses in the
radiative layer.  Therefore, the diffusion timescale must not be greater
than the cycle half-period: $t_{\rm d}\le\tau_{0.5}$.  This implies
another a theoretical constraint on the turbulent diffusivity for consistency of our dynamo model, namely $\beta_\phi\ge
L^2 / \tau_{0.5}$.  This latter constraint yields $\beta_\phi\gtrsim4\times10^{17}$~cm$^2$~s$^{-1}$.  A subtlety is that the transport
would not be a strictly isotropic diffusion but could be the result of down pumping by anisotropic 
convection \cite{Tobias:1998lr}.

One way to infer the turbulent diffusivity in evolved stars is to note
that isotopic anomalies in low-mass Red Giant Branch and AGB stars
require that material from the base of the convective zone be
transported to near the H-burning shell, processed, and returned to
the convective envelope \cite{Wasserburg:1995cj, Nollett:2003il}.
This so-called ``cool bottom processing'' (CBP) is thought to be
magnetically driven \cite{Busso:2007jw, Nordhaus:2008fk}.  In the
magnetic mixing scenario, a lower bound on the turbulent diffusivity
is $\sim$$7\times10^{15}$ cm$^2$ s$^{-1}$ \cite{Nordhaus:2008fk}.
While the actual diffusion coefficients in evolved star interiors are
unknown, the values inferred from CBP are more than an order of
magnitude lower than what are required for the envelope dynamo
scenario to transport $\sim$10$^4$-G fields all the way to the WD
surface.  Higher field strengths (in particular, $\sim$MG fields) lead
to shorter buoyant-rise times and smaller cycle half-periods, and
therefore require even greater diffusivity.  In short, a substantial macroscopic
diffusivity would be needed to transport megagauss fields to the WD surface in
AGB stars.  We presently do not have independent constraints on these coefficients.

\subsection{Disk Dynamo}
\label{ssec:DiskDynamo}
If the field generated in the envelope cannot diffuse to the WD
surface, an alternative possibility is amplification in an accretion
disk that forms when the companion is tidally disrupted inside the
common envelope \cite{Reyes-Ruiz:1999lr}.  As shown below, this
scenario is attractive because it provides a natural mechanism for
transporting the field to the proto-WD surface.

We consider disks formed from companions
spanning the range from sub-Jupiter-mass planets (see below) to brown dwarfs to low-mass stars
(i.e. $M_{\rm c}$ between $\sim$$0.1M_{\rm J}$ and a few times
$10^2M_{\rm J}$).\footnote{$M_{\rm J}$ is the mass of Jupiter.  This classification as planet or brown dwarf is based solely on mass \cite{Spiegel:2010lr}}  Tidal disruption of the companion results in formation of a disk
inside the AGB star.  This occurs at the tidal shredding
radius,
which we estimate as
$R_{\rm s} \simeq R_{\rm c} (2M / M_{\rm c})^{1/3}$,
where $r_{\rm c}$ and $M_{\rm c}$ are the radius and mass of the
companion and $M$ is the stellar mass interior to $R_{\rm s}$
\cite{Nordhaus:2010lr}.
We note that, since planetary and low-mass stellar companions to
solar-type stars are significantly more plentiful than brown dwarfs
\cite{Wright:2009fk,Marcy:2000uq,Farihi:2005qy}, disks at the upper or
lower mass range may be more common than those in the intermediate
mass range.

The disk is ionized and susceptible to the development of magnetized
turbulence (e.g.~via the magneto-rotational instability
\cite{Balbus:1998lr}).  Accretion towards the central proto-WD occurs
on a viscous timescale given by $t_{\rm visc} \simeq R^{2}/\nu \simeq P_{\rm orb} / \alpha_{\rm ss} (H / R)^2$,
where $P_{\rm orb}$ is the Keplerian orbital period at radius $R$, $\nu = \alpha_{\rm ss}c_{\rm s}H = \alpha_{\rm
  ss}(H/R)^{2}R^{2}\Omega$ is the effective kinematic viscosity, $H$
is the disk scaleheight, $c_{\rm s} = H\Omega$ is the midplane soundspeed, 
$\Omega = 2\pi / P_{\rm orb}$
is the Keplerian orbital frequency, and $\alpha_{\rm ss}$ is the
dimensionsless Shakura-Sunyaev parameter that characterizes the
efficiency of angular momentum transport \cite{Shakura:1973lr}.  The
initial accretion rate can be approximated as
\begin{eqnarray}
\dot{M} &\sim& \frac{M_{\rm c}}{t_{\rm visc}|_{r_{\rm s}}} \approx 7M_{\odot}{\,\rm yr^{-1}}\times \label{eq:mdot0} \\
&& \left(\frac{\alpha_{\rm ss}}{10^{-2}}\right)\left(\frac{M_{\rm c}}{30 M_{\rm J}}\right)^{3/2}\left(\frac{r_{\rm c}}{r_{\rm J}}\right)^{-1/2}\left(\frac{H/R}{0.5}\right)^{2} \nonumber \, ,
\end{eqnarray}
where we have scaled $r_{\rm c}$ to the radius of Jupiter\footnote{
  $r_{\rm J}$ is the radius of Jupiter.  Note that $r_{\rm c}$ depends
  only weakly on mass for the companions we consider.} and have scaled
$H/R$ to a large value $\sim 0.5$ because the disk cannot cool
efficiently at such high accretion rates and would be geometrically
thick.  Since the companions under consideration lead to disks much more dense than the stellar envelope, it is reasonable to ignore any interaction between the disk and the star.

Equation (\ref{eq:mdot0}) illustrates that for $M_{\rm c} \sim
0.1-500 M_{\rm J}$, and for typical values of $\alpha_{\rm ss}
\sim 0.01-0.1$, $\dot{M}$ is $\sim$3 to 9 orders of magnitude larger
than the Eddington accretion rate of the proto-WD
(i.e.,
$\dot{M}_{\rm Edd} \sim 10^{-5} M_{\odot}$~yr$^{-1}$).  At first, it
might seem apparent that inflow of mass onto the WD surface would be
inhibited by radiation pressure in such a scenario.  However, this
neglects the fact that, at sufficiently high accretion rates, photons
are trapped and advected to small radii faster than they can diffuse
out \cite{Colgate:1971fk, Katz:1977qy, Blondin:1986uq,
  Chevalier:1989fj, Houck:1992kx}.  In this `hyper-critical' regime,
accretion is possible even when $\dot{M} \gg \dot{M}_{\rm Edd}$.

We evaluate the possibility of hyper-critical accretion by estimating
the radius interior to which the inward accretion timescale, $t_{\rm
  visc}$, is less than the timescale for photons to diffuse out of the
disk midplane, $t_{\rm diff}$ (which is approximately $H\tau/c$, where
$\tau$ is the vertical optical depth
\cite{Katz:1977qy, Begelman:1979yq, Blondin:1986uq}).  This ``trapping
radius''
is given by $R_{\rm tr} = \dot{M}\kappa H / 4\pi R c$ where $\kappa$ is the opacity.
An important quantity is the ratio of the trapping radius to the outer
disk radius $R_{\rm s}$ (coincident with the tidal shredding radius) :
\begin{eqnarray}
\frac{R_{\rm tr}}{R_{\rm s}} &\approx& 1.0\times 10^{4}\left(\frac{\alpha_{\rm ss}}{10^{-2}}\right)\left(\frac{\kappa}{\kappa_{\rm es}}\right)\times\label{eq:trap_ratio} \\
&& \left(\frac{M_{\rm c}}{30 M_{\rm J}}\right)^{11/6}\left(\frac{M_{\rm WD}}{0.6M_{\odot}}\right)^{-1/3}\left(\frac{r_{\rm c}}{r_{\rm J}}\right)^{-3/2}\left(\frac{H/R}{0.5}\right)^{3}, \, \nonumber
\end{eqnarray}
where we have used equation (\ref{eq:mdot0}) and scaled $\kappa$ to
the electron scattering opacity $\kappa_{\rm es} = 0.4$ cm$^{2}$ g$^{-1}$.  If $R_{\rm s}<R_{\rm tr}$, then $t_{\rm visc}<t_{\rm diff}$ and photons are advected with the matter.

From equation (\ref{eq:trap_ratio}) we conclude that $R_{\rm tr}> R_{\rm s}$ for $M_{\rm c} \gtrsim 0.1M_{\rm J}$.  This implies that photon pressure will be unable to halt accretion initially and that the local energy released by accretion must be removed by advection \cite{Spruit:2001lr}.  Advection acts like a conveyor belt, nominally carrying the gas to small radii as its angular momentum is removed.  If, however, there is no sink for the hot gas, this conveyor may ``jam".  This is an important distinction between white dwarfs and neutron stars or black holes, as the latter two can remove the thermal energy by neutrino cooling or advection through the event horizon, respectively.  In contrast to a WD, neutrino cooling is ineffective and thermal pressure builds, such that radiation pressure may again become dynamically significant \cite{Houck:1992kx}.  This in turn shuts down the accretion to at most, the Eddington rate.  

\begin{figure}
\begin{center}
\includegraphics[width=8.9cm,angle=0,clip=true]{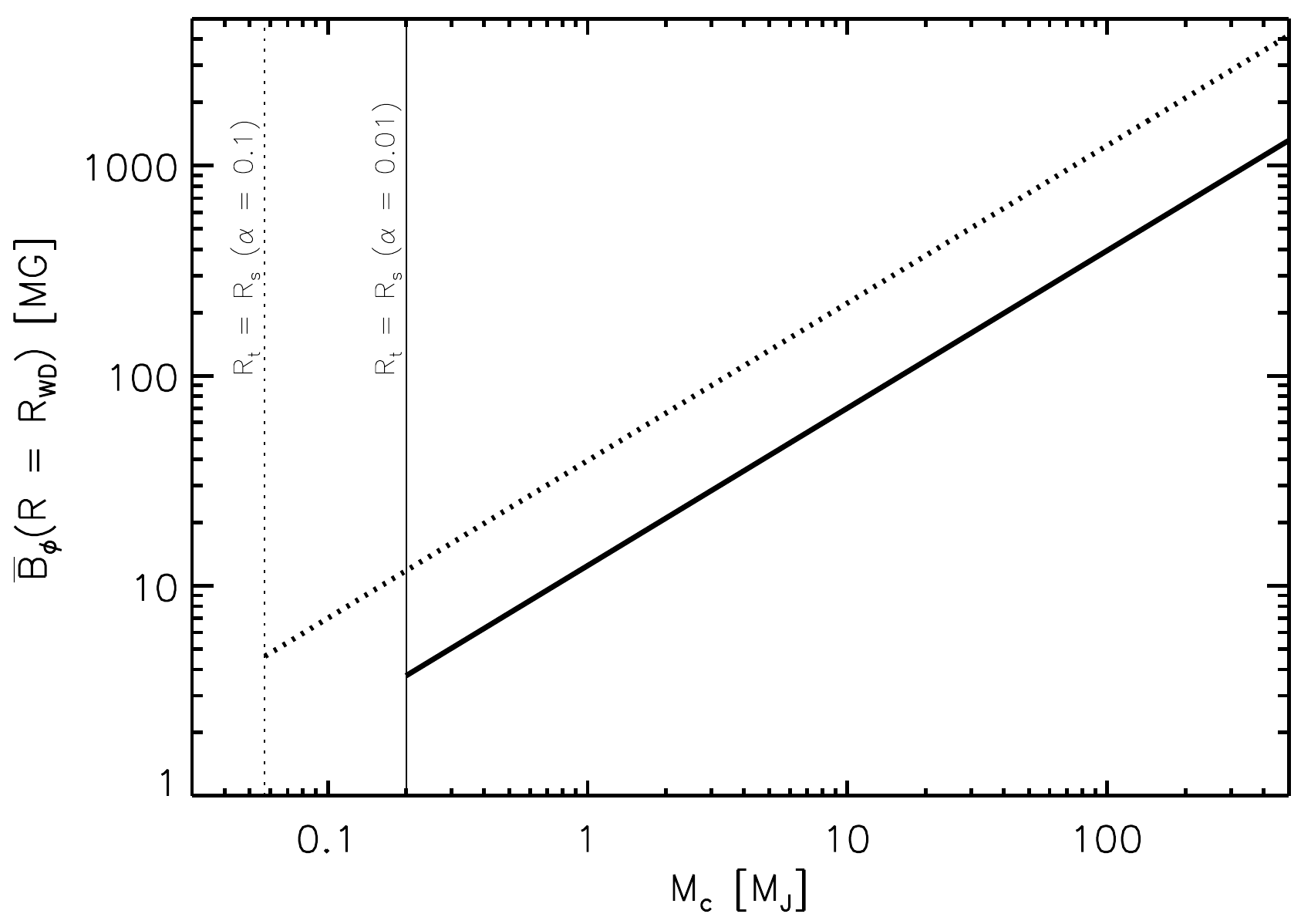}
\caption{Toroidal magnetic field strength,$\overline{B}_{\phi}$ (eq.~[\ref{eq::toroidal}]), at the WD surface
   as a function of
  the mass of the tidally disrupted companion $M_{\rm c}$.  Toroidal field strengths are presented
  for two values of the viscosity, $\alpha_{\rm ss} = 0.01$ ({\it
    solid line}) and $\alpha_{\rm ss} = 0.1$ ({\it dotted line}) and
  assuming an accretion efficiency $\eta_{\rm acc} = 0.1$.  The
  white dwarf mass and radius are $M_{\rm WD} = 0.6M_{\odot}$ and
  $r_{\rm WD} = 10^{9}$~cm, respectively.  The vertical lines show the
  companion mass above which photons are trapped in the accretion flow
   (i.e., $r_{\rm tr} > r_{\rm s}$;
  eq.~[\ref{eq:trap_ratio}]), such that super-Eddington accretion
  occurs.
\label{fig:fig1}}
\end{center}
\end{figure}

However, this scenario neglects the possibility of outflows\footnote{It's interesting to note that the presence of such outflows in the present context is coincident with the prevalence of bipolar proto-planetary nebulae.}, which can sustain inflow by carrying away the majority of the thermal energy, thereby allowing a smaller fraction of the material to accrete at a higher than Eddington rate.  Though more work is needed to assess the efficiency of outflows in the present context, radiatively-inefficient accretion flows are widely thought to be prone to powerful outflows \cite{Narayan:1995vn, Blandford:1999rt, Armitage:2000ys,Spruit:2001lr,Hawley:2002fr,Ohsuga:2005fk}.  Even if accretion occurs at, or near, the Eddington rate, the fields produced ($\gtrsim10$~MG) are still strong enough to explain the bulk of the magnetized WD population.  The origin of the strongest field systems ($\sim$100-1000 MG) may be problematic if accretion is limited to the Eddington rate.

Nevertheless, the material deposited outside the WD will be hot and virialized, forming an extended envelope with a lengthscale comparable to the radius.  Though initially hot, this material will eventually cool (on longer, stellar timescales) and become incorporated into the stellar layers near the proto-WD surface.  If this material cools at some fraction of the local Eddington luminosity, it will be incorporated onto the WD surface on a timescale $\sim$$10^2-10^4$ years, which is much less than typical AGB lifetimes.

A final complication may arise because the composition of the companion is likely to be hydrogen-rich and will begin to burn thermonuclearly as material accretes or after it settles on the proto-WD surface.  Because the energy released in burning hydrogen to helium ($\sim 7$ MeV nucleon$^{-1}$) vastly exceeds the gravitational binding energy per nucleon at the WD surface ($\sim 0.1$ MeV nucleon$^{-1}$), burning that occurs explosively could have a dynamical impact.  For the accretion rates of relevance, we estimate that the midplane temperature of the disk exceeds $\sim 10^{8}$ K, such that hydrogen will burn via the hot CNO cycle.  The hot CNO cycle occurs at a rate that is independent of temperature and is itself thermally stable.  However, hydrogen-rich material deposited in the He burning layer may, under some circumstances, trigger a thermonuclear runaway \cite{Podsiadlowski:2010lr}, which could remove the accreted mass from the proto-WD surface, or even eject the common envelope entirely.  In what follows we assume stable burning and set aside this important caveat.  Additional work will, however, be required to address what conditions are required for stable versus explosive burning.

Due to the presence of shear in the disk,\footnote{For black hole disks, differential rotation near the disk midplane is preserved even in a thick-disk, super-Eddington context \cite{Qian:2009uq}.  However, for white dwarf disks, photons are not advected into, and lost to, the black hole.} the MRI is a likely source of turbulence,
which in turn amplifies magnetic field.  Large-scale fields produced by the MRI
have been modeled via $\alpha-\Omega$ dynamos at various levels of sophistication (see \cite{Gressel:2010fj} for the most recent example).  However, for the purposes of estimating orders of magnitude of the fields, approximate values from less sophisticated treatments can be employed.  Although this situation is qualitatively similar to that
described in the previous subsection,
the dynamo we envision here operates in the accretion disk and not the
envelope.  The Alfv\'{e}n velocity in the disk obeys $v_{\rm
  a}^2=\alpha_{\rm ss} c_{\rm s}^2$ \cite{Blackman:2001uq}, such that
the mean toroidal field at radius $R$ is given by
\cite{Blackman:2001uq}
\begin{eqnarray}
\label{eq::toroidal}
\overline{B}_\phi &\sim& \left(\frac{\dot{M}\Omega}{R}\frac{R}{H}\right)^{1/2}  \\
&\approx& 160~\rm{MG}\left(\frac{\eta_{\rm acc}}{0.1}\right)^{1/2}\left(\frac{\alpha_{\rm ss}}{10^{-2}}\right)^{1/2}\left(\frac{M_{\rm c}}{30 M_{\rm J}}\right)^{3/4}\times \nonumber \\
&&\left(\frac{M_{\rm WD}}{0.6M_{\odot}}\right)^{1/4}\left(\frac{r_{\rm c}}{r_{\rm J}}\right)^{-1/4}\left(\frac{H/R}{0.5}\right)^{1/2}\left(\frac{R}{10^{9}\rm cm}\right)^{-3/4} \, , \nonumber
\end{eqnarray}
where in the second equality we have substituted equation
(\ref{eq:mdot0}) for $\dot{M}$, and
multiplied by the factor $0.1 \lesssim \eta_{\rm acc} \le 1$ to
account for the possibility of outflows as described above.  If an
$\alpha-\Omega$ dynamo operates, the mean poloidal field
$\overline{B}_p$ is related to the toroidal field via
$\overline{B}_p=\alpha_{\rm ss}^{1/2}\overline{B}_\phi$
\cite{Blackman:2001uq}.  However, regardless of whether a large scale
field is generated, a turbulent field of magnitude
$\overline{B}_{\phi}$ is likely to be present and contributing significantly to the Maxwell stresses
responsible for disk accretion.

Figure~\ref{fig:fig1} shows the toroidal field evaluated near the WD
surface $R \approx R_{\rm WD} \approx 10^{9}$~cm as a function of
companion mass $M_{\rm c}$, calculated for two different values of the
viscosity ($\alpha_{\rm ss} = 0.01$ and $0.1$).  In both cases, we assume that $\eta_{\rm acc} = 0.1$.  Note that
for the range of relevant companion masses $M_{\rm c} \sim 0.1 - 500$
$M_{\rm J}$, $\overline{B}_{\phi} \sim 10-1000$~MG, in precisely the
correct range to explain the inferred surface field strengths of
HFMWDs.  

Although our results suggest that companion accretion can produce the field strengths necessary to explain HFMWDs, for the field to be present on the surface of the WD when it forms, it must at a minimum survive the remaining lifetime of the star.  In particular, the field will decay due to ohmic diffusion on the timescale
\begin{eqnarray}
&&\tau_{\rm decay} \sim \frac{3(\Delta R)^{2}}{\eta_{\rm s}} \sim 4\times 10^{6}{\,\rm yr}\left(\frac{T}{10^{8}\rm K}\right)^{3/2}\times \nonumber \\
&&\left(\frac{{\rm ln}\Lambda}{10}\right)^{-1}\left(\frac{R_{\rm WD}}{10^{9}{\,\rm cm}}\right)^{2}\left(\frac{M_{\rm WD}}{0.6M_{\odot}}\right)^{-2}\left(\frac{\eta_{\rm acc}}{0.1}\right)^{2}\left(\frac{M_{\rm c}}{30M_{\rm J}}\right)^{2}, \nonumber \\
\label{eq:taudecay}
\end{eqnarray}
where $\eta_{\rm s} \approx 5\times 10^{12}({\rm ln}\Lambda/10)T^{3/2}$ cm$^{2}$s$^{-1}$ is the Spitzer resistivity, ln$\Lambda$ is the Coulomb logarithm, $T$ is the stellar temperature near the proto-WD surface, and $\Delta R = \eta_{\rm acc}M_{\rm c}/4\pi r_{\rm WD}^{2}\bar{\rho}$ is the final thickness of the accreted companion mass after it is incorporated into the surface layers of proto-WD, where $\bar{\rho} \equiv M_{\rm WD}/(4\pi r_{\rm WD}^{3}/3)$ is the mean density of the proto-WD.

Equation (\ref{eq:taudecay}) shows that for typical core temperatures during the AGB phase $T \sim 10^{8}$ K, $\tau_{\rm decay}$ exceeds the mean AGB lifetime $\tau_{\rm AGB} \sim$ Myr for $M_{\rm c} \gtrsim 20M_{\rm J}$.  This suggests that fields may survive Ohmic decay, at least for very massive companions.  Furthermore, the magnetized companion material may become incorporated into the degenerate WD core long before the AGB phase ends, in which case the much higher conductivity due to degenerate electrons substantially increase the Ohmic decay timescale over that given by equation (\ref{eq:taudecay}), thereby ensuring long-term field survival.

\section{Conclusions}
\label{sec::Conclusions}
Recent observational evidence from the Sloan Digital Sky Survey strongly suggests that
high-field magnetic white dwarfs originate from binary interactions.
In particular, it was proposed that the progenitors of HFMWDs are
binary systems that
evolve through a common envelope phase \cite{Tout:2008qy}.  In this
scenario, companions that
survive CE evolution (leaving tight binaries) may produce magnetic
cataclysmic variables, while those that do not survive (i.e. that
merge) may produce isolated HFMWDs.

To investigate this hypothesis, we have estimated the magnetic fields
resulting from a low-mass companion embedded in a CE phase with a
post-MS giant.  During in-spiral, the companion transfers energy and
angular momentum to the envelope.  The resulting shearing profile,
coupled with the presence of convection, can amplify large-scale
magnetic fields via dynamo action in the envelope.  We incorporate the
back-reaction of the magnetic field growth on the shear, which results
in a transient dynamo.  The fields generated at the interface between
the convective and radiative zones are weak ($\sim$2$\times10^4$~G) and would have
to diffuse to the WD surface and anchor there if they are to explain the HFMWDs.  A successful envelope dynamo scenario must also survive to the end of the AGB phase, which probably means only if the companion ejects the turbulent envelope; the dynamo is transient as long as there is a source of turbulent diffusion.  The bulk of the systems formed this way might
be post-CE, tight binaries -- potentially magnetic CVs.

To explain isolated HFMWDs, an alternative to a dynamo operating in
the envelope is a dynamo operating in an accretion disk.  During the
CE phase, if the companion is of sufficiently low mass, it avoids
prematurely ejecting the envelope.  Instead, in-spiral proceeds until the
companion is tidally shredded by the gravitational field of the
proto-WD.  The subsequent formation of an accretion disk, which also
posesses turbulence and shear, can amplify magnetic fields via dynamo
action.  For the range of disrupted companions considered here, the disk initially accretes at super-Eddington values.  In this hyper-critical regime, the timescale for photon diffusion out of the disk is longer than the viscous timescale. For a range of disrupted-companion disks, we find that the
saturated toroidal mean-field attains values between a few and a few thousand megagauss.  Amplification of the magnetic field in a super-Eddington
accretion disk is attractive as it reproduces the observed range of HFMWDs and naturally transports magnetic flux to the WD surface.

High-mass stars may also undergo common envelope interactions in the
presence of close companions.  If the common envelope field mechanisms
described here operate in high-mass stars, then the result could
be strong field neutron stars or magnetars (neutron stars with
magnetic fields in excess of $\sim$$10^{14}-10^{15}$~G).  In
particular, formation of an accretion disk from an engulfed companion
during a red supergiant phase could produce a magnetized WD core.  In
the eventual core-collapse and stellar supernova explosion (possibly
driven by the neutrino mechanism; \cite{Nordhaus:2010kx}), the
magnetized WD core collapses to a neutron star.  If simple flux
freezing operates (itself an open question) and the initial magnetized
core is on the order of $\sim$100-1000~MG, homologous collapse to a
neutron star would generate $\sim$$10^{14}-10^{15}$~G fields.  Typical
neutron stars that possess modest field strengths may originate from
core-collapse supernova of single stars or stars without having
incurred a CE phase.  Note that what we are proposing here is an alternative to the neutrino-driven convection dynamo described in \cite{Duncan:1992kx,Thompson:1993yq}.  In our model, the engulfment of a companion and the formation of an accretion disk naturally provides fast rotation, magnetized turbulence and differential rotation.  We emphasize that the viability of this mechanism depends on the length scale of the magnetic field deposited in the pre-collapse core, which must be sufficiently large to produce, upon collapse, the dipole-scale, volume-encompassing fields necessary to account for the spin-down behavior of magnetars and the energy budget of their giant flares \cite{Woods:2006qy}.

In summary, common envelope evolution as the origin of
strongly-magnetized, compact objects seems plausible.
Whether this hypothesis is ultimately found to be viable will depend
on the statistics of low-mass stellar and substellar companions to
stars of similar masses to (or somewhat higher masses than) the Sun.
The numerous radial velocity searches of the last 20 years
have revealed a number of such companions \cite{Duquennoy:1991fk,Halbwachs:2003qy,Raghavan:2010lr}.  The precise occurrence rate of
companions in orbits that could lead to the kind of disk-dynamo mechanism
described above remains unclear (though if such companions turn out to
be rarer than the HFMWD fraction, then the SDSS data
indicating a binary origin would be puzzling).  Further theoretical
work into the binary origin of HFMWDs and magnetars requires the
development of multi-dimensional, magnetohydrodynamic simulations of
the CE phase.  Such an approach has already had initial success for purely hydrodynamic
adaptive mesh refinement simulations
\cite{Ricker:2008fk}.

\begin{acknowledgments}
We thank Alberto Lopez, Jay Farihi, Jim Stone, Jeremy Goodman, Kristen Menou, Deepak Raghavan, Andrei
Mesinger, Adam Burrows and Fergal Mullally for thoughtful discussions.  JN
acknowledges support for this work from NASA grant HST-AR-12146.  DSS
acknowledges support from NASA grant NNX07AG80G.  Support for BDM is
provided by NASA through Einstein Postdoctoral Fellowship grant number
PF9-00065 awarded by the Chandra X-ray Center, which is operated by
the Smithsonian Astrophysical Observatory for NASA under contract
NAS8-03060.  EGB acknowledges support from grants NSF PHY-0903797 and
NSF AST-0807363.
\end{acknowledgments}



\end{article}

\end{document}